\documentclass[conference]{IEEEtran}
\IEEEoverridecommandlockouts
\usepackage[T1]{fontenc}
\usepackage{cite}
\usepackage{amsmath,amssymb,amsfonts, amsthm}
\usepackage{mathptmx}
\usepackage{algorithmic}
\usepackage{graphicx}
\usepackage{textcomp}
\usepackage{braket}
\usepackage{qcircuit}
\usepackage{xcolor}
\usepackage{listings}
\usepackage{todonotes}
\usepackage{dsrm}
\usepackage{pifont}
\usepackage{microtype}
\usepackage{booktabs}
\usepackage{listliketab}
\usepackage{hyperref}
\usepackage{color}
\definecolor{dkgreen}{rgb}{0,0.6,0}
\definecolor{gray}{rgb}{0.5,0.5,0.5}
\definecolor{mauve}{rgb}{0.58,0,0.82}

\ifCLASSOPTIONcompsoc
    \usepackage[caption=false, font=normalsize, labelfont=sf, textfont=sf]{subfig}
\else
\usepackage[caption=false, font=footnotesize]{subfig}
\fi

\newcommand{\xmark}{}
\newtheorem{definition}{Definition}

\makeatletter
\def\fps@figure{htbp}
\def\fps@table{htbp}
\makeatother

\newcounter{criterion}
\newcommand{\criterion}[1]{%
  \refstepcounter{criterion}%
  \indent
  \textit{\thecriterion) #1:}%
}

\begin{document}

\title{Productive Quantum Programming Needs Better Abstract Machines}

\author{
\IEEEauthorblockN{Santiago Núñez-Corrales\IEEEauthorrefmark{1}, Olivia Di Matteo\IEEEauthorrefmark{2}, John Dumbell\IEEEauthorrefmark{3}, Marcus Edwards\IEEEauthorrefmark{2}\IEEEauthorrefmark{4}, Edoardo Giusto\IEEEauthorrefmark{5},\\ Scott Pakin\IEEEauthorrefmark{6}, Vlad Stirbu\IEEEauthorrefmark{7}}
\IEEEauthorblockA{\IEEEauthorrefmark{1} \textit{NCSA/IQUIST}, \textit{University of Illinois Urbana-Champaign}, Urbana IL, USA\\
nunezco2@illinois.edu}
\IEEEauthorblockA{\IEEEauthorrefmark{2} \textit{Electrical and Computer Engineering and Stewart Blusson Quantum Matter Institute,}\\ \textit{The University of British Columbia}, Vancouver BC, Canada\\ \{olivia, msedwards\}@ece.ubc.ca}
\IEEEauthorblockA{\IEEEauthorrefmark{3}\textit{Quantum Science \& Exploratory Research}, \textit{Oxford Quantum Circuits}, Reading, United Kingdom\\jdumbell@oxfordquantumcircuits.com}
\IEEEauthorblockA{\IEEEauthorrefmark{5}\textit{DIETI}, \textit{University of Naples, Federico II}, Naples, Italy\\
egiusto@ieee.org}
\IEEEauthorblockA{\IEEEauthorrefmark{6}\textit{Computer, Computational, and Statistical Sciences Division}, \textit{Los Alamos National Laboratory}, Los Alamos, NM, USA\\
pakin@lanl.gov}
\IEEEauthorblockA{\IEEEauthorrefmark{7}\textit{Faculty of Information Technology, University of Jyväskylä}, Jyväskylä, Finland\\
vlad.a.stirbu@jyu.fi}
}

\maketitle
\thispagestyle{plain}
\pagestyle{plain}

\begin{abstract}
An effective, accessible abstraction hierarchy has made using and programming computers possible for people across all disciplines.  Establishing such a hierarchy for quantum programming is an outstanding challenge, especially due to a proliferation of different conventions and the rapid pace of innovation.  One critical portion of the hierarchy is the \emph{abstract machine}, the layer that separates a programmer's mental model of the hardware from its physical realization. Drawing on historical parallels in classical computing, we explain why having the ``right'' quantum abstract machine (QAM) is essential for making progress in the field and propose a novel framework for evaluating QAMs based on a set of desirable criteria. These criteria capture aspects of a QAM such as universality, compactness, expressiveness, and composability, which aid in the representation of quantum programs. By defining this framework we take steps toward defining an optimal QAM\@. We further apply our framework to survey the landscape of existing proposals, draw comparisons, and assess them based on our criteria. While these proposals share many common strengths, we find that each falls short of our ideal.  Our framework and our findings set a direction for subsequent efforts to define a future QAM that is both straightforward to map to a variety of quantum computers, and provides a stable abstraction for quantum software development.
\end{abstract}


\section{Introduction}

Quantum programming remains largely a difficult task~\cite{di2024abstraction}. Several factors contribute to the present state of affairs. First, due to their macroscopic, classical state of being, humans lack direct phenomenological experience with quantum effects. Second, quantum hardware is reaching scales at which it makes sense to think of them as computer systems rather than as physics experiments. Finally, quantum software is now emerging as its own ecosystem. Naturally, quantum programming languages reflect the state of evolution of this burgeoning field: programs currently relate more to quantum device physics than directly to expressing solutions to computational problems.

As a consequence, \emph{productive} quantum programming is still unattainable for most people without physics-based research training.  Without a detailed understanding of quantum hardware, programmers are frustrated by the effort needed for algorithms discovery and performance optimization when using the prevailing circuit model.  Today's quantum programming models require users to juggle a large number of concerns, reflecting poorly managed complexity and low separation of concerns.

Abstraction hierarchies are essential for managing complexity and providing expressiveness.  Cognitive science research indicates that the more simultaneous details required for a particular task, the lower the subject's ability to remember them while performing that task~\cite{miller1956magical}. Limiting the number of facts to remember has been historically a role for the \textit{execution model}, often fulfilled by an \textit{abstract machine}. An abstract machine is a mathematical specification of a computer implementable by harnessing known physical laws. It presents a convenient standard interface to programmers, so they can avoid the underlying hardware details.

The success of von Neumann's random access machine~\cite{von1993first} in becoming the predominant blueprint for instruction set architectures (ISAs)  in classical computing is largely explained by its affinity with how programmers think about problems, rather than to how computing hardware actually works. In a very pragmatic sense, ``\emph{programs are meant to be read by humans and only incidentally meant for computers to execute}''~\cite{abelson1996structure}. Consequently, our intent here is to characterize the state of quantum abstract machines and determine whether existing ones contain powerful enough primitive expressions, means of combination, and means of abstraction to enable the quantum revolution we all anticipate. To this end, we put forward a novel framework for assessing quantum abstract machines, using criteria pertaining to the representation of quantum programs. We believe this framework to be a meaningful step towards development of an optimal QAM, as it outlines the characteristics one ought to have, and corresponding outcomes for the programmers, physicists, and engineers who use it.

\section{Background and motivation}

\subsection{Abstract machines' influence in computing}

The need for better quantum programming abstractions and program representations, combined with a recently proposed programming hierarchy~\cite{di2024abstraction}, forms the driving question from the discussion above: what must a quantum abstract machine encompass to simultaneously (1)~facilitate reasoning about a program's behavior, (2)~abstract away low-level hardware details, and (3)~enable programmers to ``acquire good algorithms and idioms''?~\cite{abelson1996structure}. All three stated requirements can be satisfied by an appropriate abstract machine model.

\begin{definition}
An \emph{abstract machine} is a model of computation that specifies what data can be input, output, and stored, and provides a semantics of operations that can be performed on that data.
\end{definition}

This paper considers abstract machines capable of general problem solving in the sense of Turing-computable functions~\cite{gherardi2011alan}, or functions whose solution can be achieved through a program of finite size and after a finite number of steps.  We are particularly interested in abstract machines that facilitate problem-solving and algorithm expression---reducing effort from both humans and compilers---by supporting the design and implementation of high-level programming languages. Productive programming requires mental models of computation powerful enough to solve a wide variety of problems yet lean enough to fit (cognitively) in one's mind. Hence, it is worth differentiating between \emph{expressiveness} in the formal sense (i.e.,~the number of classes of problem that can be solved) from its more pragmatic interpretation as how clearly and concisely programmers can express their intent through code.

Abstract machines execute programs composed of viable operations from the represented model, commonly in the form of instructions. Instructions are tightly linked to the abstract machine's semantics and can take many forms but largely represent a transition in the underlying physical machine's state.  Hence, a classical instruction may access or store memory in registers, perform arithmetic, or branch to another instruction.  In contrast, a quantum instruction may execute a pulse sequence, perform readout, or take a measurement. While evident differences exist between classical and quantum hardware, they are similar at a formal level of analysis, which justifies our reasoning and historical reconstruction below.

\subsubsection{Historical Abstract Machines}

General problem solving can be expressed by multiple, qualitatively different abstract machine models, so choosing an appropriate model is dictated by desired constraints and use cases. Since both classical and quantum abstract machines operate in the class of recursively enumerable problems, it is worth revisiting the kinds of questions answerable with two canonical abstract machines in classical computing: Turing machines and electronic circuits.

Turing machines (TMs) were mostly useful to define computability. They are extremely limited in practice, as their expressiveness is too poor to be usable for concrete problem solving beyond the theory of computation, and too impractical to implement in actual hardware. The set of possible operations of a TM and the model of information storage lack flexibility and convenience. In the more constructive view Church took on computability~\cite{church1932set} functions are intuitive, but their use is as a meta-language focused on teasing out logic constraints rather than prescribing usable constructs.

More practically, Bush's observation that electronic circuits could be used to implement Boolean logic~\cite{shannon1938symbolic} was a fortunate match between bi-stable (binary) systems with sufficient reliability and finite models of arithmetic whose operations exhibit properties which implement useful approximations to the algebra of real fields. This match was crucial, albeit circumstantial: known abstract machines filtered existing technologies depending on implementability and convenience. Circuits do provide a better interface to build programs than exact physical reproductions of Turing machines or lambda calculus, but they remain far from being intellectually efficient when programs need to scale. Specifying circuits became an exercise in modularity with the intent of obtaining hardware organization destined to last much longer than software.

\subsubsection{Modern Abstract Machines}

What, then, enabled the community to rise above circuits in modern classical computing? The EDVAC report~\cite{von1993first} and its subsequent analysis~\cite{burks1946preliminary} contain the first descriptions of \emph{instructions}, as we know them today.

\begin{definition}
An \emph{instruction} is an atomic, composable procedural abstraction.
\end{definition}

An \textit{atomic} instruction has a recognizable identity, and one can reason about its effects in terms of reachable program states without having to leave the same level of abstraction. It is \textit{composable} if its presence in the execution of a program interacts with other instructions in a non-trivial way through \textit{program state}. Finally, it is considered \textit{procedural} because the order in which instructions appear matters, and differences in computation execution paths can be explained and reasoned symbolically in terms of differences in instruction sequencing, value-dependent choices, or repetition~\cite{jackson2002jsp}.

The representation and operation of instructions in an abstract machine must be predictable and regimented, including the way they transform machine state, represent data, and interact with each other. This definition, and the specific rules by which the instructions interact with the abstract machine and each other, are called its \emph{interface} or \emph{contract}. The manifestation of this interface is the Instruction Set Architecture (ISA).

\begin{definition}
An \emph{instruction set architecture} concretizes the abstraction provided by an abstract machine into a collection of discrete, mechanizable operations. 
\end{definition}

Instructions separate concerns across multiple dimensions. The first dimension is separating information content from hardware control in specifiable transformations: we can concentrate on writing programs using instructions satisfying a contract without worrying about finely granular hardware events and signals. 
The second dimension is the separation between information storage and information interpretation.  When reasoning about the effects of a program, a programmer works at the level of formal statements (mathematics), translates statements into programs (instructions), and ideally uses knowledge of the limitations of finite representations to validate program correctness (proofs). Programmers no longer need to worry about the fate of individual bits or the sequence of gates the machine executes.

Abstract machines heavily explain the rise of high-level languages, a fact traceable down to four key events after the implementation of EDVAC. First, Asser in 1959~\cite{asser1959turing} proved the equivalence between a machine with random memory access and universal Turing machines as a byproduct of studying their predictability through Markov models. Kaphengst~\cite{kaphengst1959abstrakte} streamlined the resulting machine as an interesting construction by itself capable of arithmetic. Backus and collaborators on ALGOL 60 defined the entry point for programming language research by showing that deliberate syntax design matters for the convenience of programmers~\cite{backus1959syntax,backus1960report}. Elgot and Robinson~\cite{elgot1964random} showed the convergence of both lines of research into random-access stored program abstract machines, mathematical devices that could be fleshed out through hardware renditions and at the same time programmed at a different level than that dictated by their internal organization.

Both theory and practice across an integrated community were required to produce abstract computing machines, unknowingly ensuring that programs could later sit on a more symbolic and predictable ground. Classical computation being compliant with classical mechanics --laws that align well with human intuitions-- appears in hindsight to have accelerated this process. We are now in a position to critique the current state of quantum abstract machines, learn from history --both quantum and classical-- and define what we see as being needed by a future quantum abstract machine.

\subsection{Existing QAM proposals}

\begin{figure*}
     \centering
     \subfloat[\label{fig:qtm}]{\includegraphics[trim={0 0.35cm 0 0.35cm},width=0.5\linewidth]{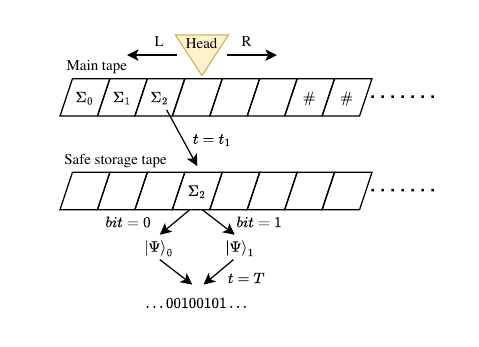}}
     \hfill
     \subfloat[\label{fig:qram}]{\includegraphics[trim={0 0.35cm 0 0.35cm},width=0.5\linewidth]{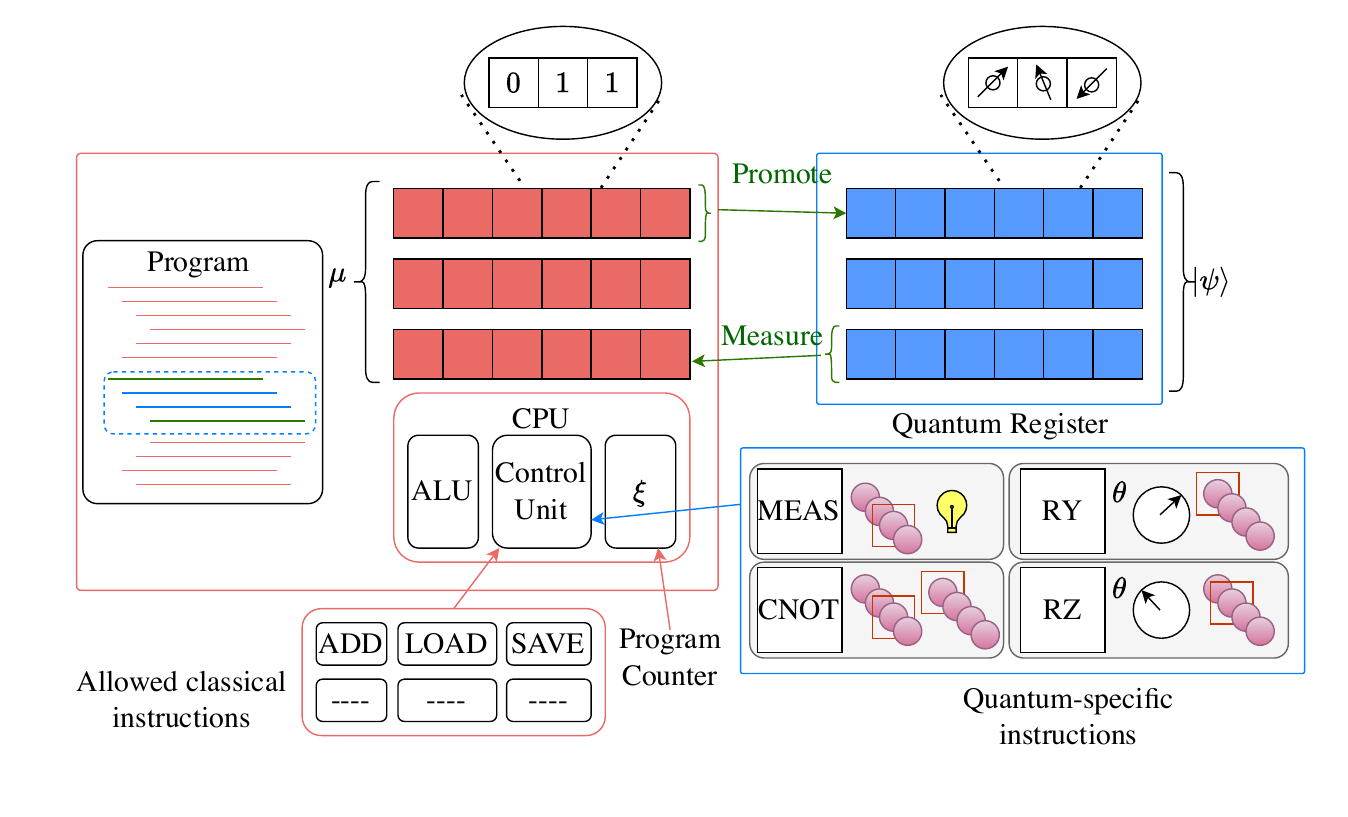}}
     \\
     \subfloat[\label{fig:qlc}]{\includegraphics[trim={0 0.35cm 0 0.35cm},width=0.5\linewidth]{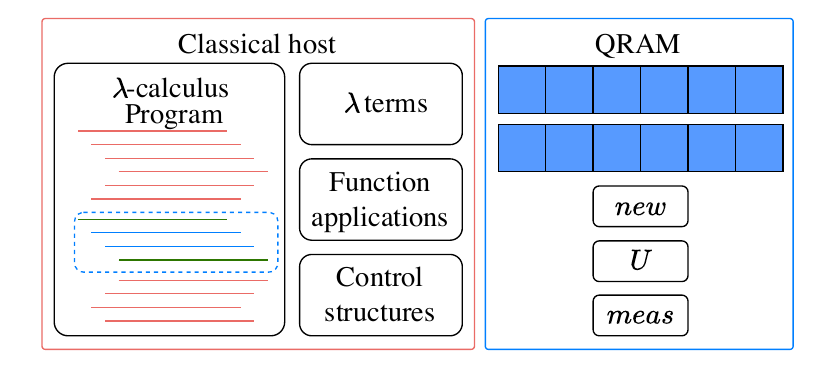}}
     \hfill
     \subfloat[\label{fig:qrasp}]{\includegraphics[trim={0 0.35cm 0 0.35cm},width=0.5\linewidth]{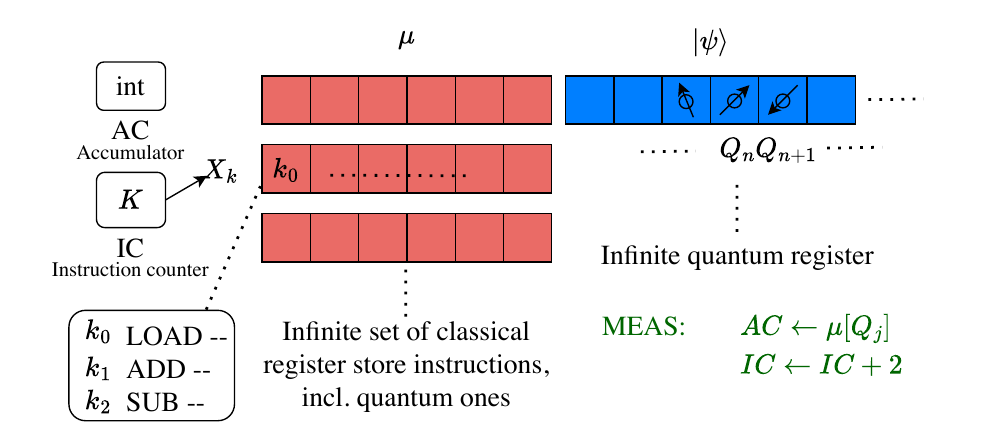}}
     \\
     \subfloat[\label{fig:qrm}]{\includegraphics[trim={0 0.35cm 0 0.35cm},width=0.5\linewidth]{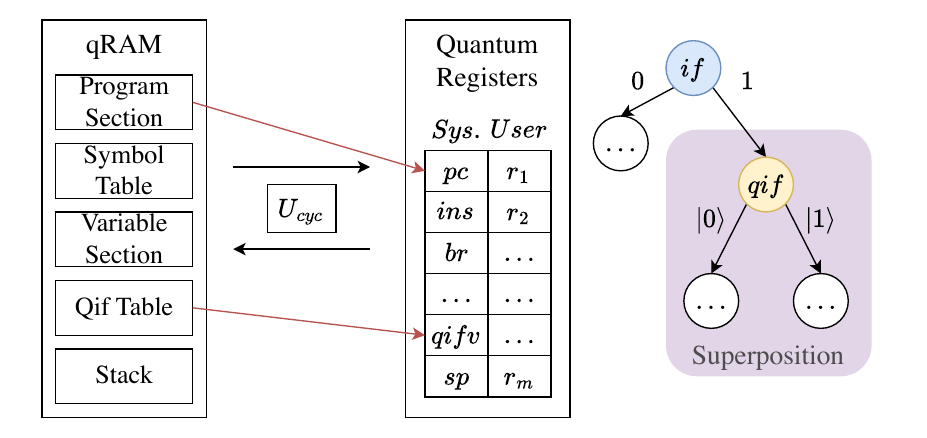}}
     \hfill
     \subfloat[\label{fig:qcm}]{\includegraphics[trim={0 0.35cm 0 0.35cm},width=0.5\linewidth]{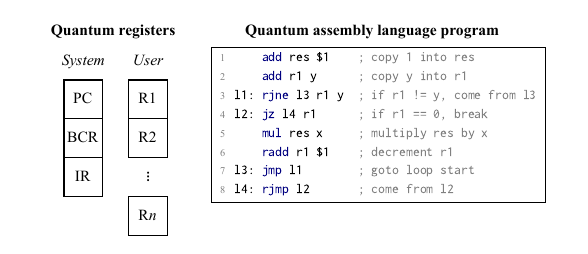}}
     \caption{Graphical representations of: (a)~the Quantum Turing Machine~\cite{bernstein1993quantum}, (b)~the Quantum Random Access Machine~\cite{knill1996conventions,miszczak2012random}, (c)~the Quantum Lambda Calculus Machine\cite{selinger2009quantum}, (d)~the Quantum Random Access Stored Program Machine\cite{wang2023quantum}, (e)~the Quantum Register Machine\cite{zhang2024qrm}, and (f)~the Quantum Control Machine (with the code example taken verbatim from the QCM paper)~\cite{yuan2024qcm}}
     \label{fig:qms}
 \end{figure*}

We consider six abstract machine proposals in this work: the quantum Turing machine, quantum lambda calculus, quantum random-access machine, quantum random-access stored program, quantum register machine, and quantum control machine. Visual representations of these machines are provided in Fig.~\ref{fig:qms}.


The first quantum abstract machine considered---and the earliest to be published---is a universal \emph{quantum Turing machine} (QTM).
Several definitions of the QTM have been proposed~\cite{deutsch1985quantum,deutsch1989qcnetworks,bernstein1993quantum,wang2023quantum}, each with its own semantics.  In this paper, we focus on the definition given by Bernstein and Vazirani~\cite{bernstein1993quantum} because it is the one used by DiVincenzo to prove two-qubit gate-set universality~\cite{divincenzo1995two}.  This QTM supports both superpositions of classical configurations and unitary evolution.
The state of a QTM is a linear superposition of classical configurations, and the QTM's dynamics are governed by unitary transformations.
A QTM consists of a finite set of quantum states $Q$, a finite classical alphabet $\Sigma$, and a
quantum finite state control
$\delta: Q \times \Sigma \times \Sigma \times Q \times \{L, R\} \rightarrow \mathbb{C}$
where $\delta(p,\sigma,\tau,q,d)$
gives the amplitude with which the machine in state $p$ reading a $\sigma$ will write a $\tau$, enter state
$q$, and move in direction $d$.
This transition function specifies a linear mapping $M_\delta$ (the time evolution operator) in the infinite dimensional space of superpositions of configurations.
A QTM is said to be \emph{well-formed} if its time evolution operator is unitary.
An \textit{observation} of some bit of the superposition of a QTM returns a zero (one) with probability according to the sum of squared magnitudes of
configurations in the superposition with a zero (one).
The QTM observation model supports observation only once at the end of the computation.
Any intermediate observation can be simulated by copying the relevant bit to a protected area of the tape.
This partitions the quantum state into non-interfering subspaces (e.g., for bit 0 and bit 1), which evolve independently.
Later observation of the saved bit yields the same outcome distribution as if it had been observed earlier.
Thus, multiple observations effectively can be deferred to a single time step without altering the computation's statistical behavior.

Two aspects of the QTM are relevant for our discussion. First, while the state space is finite-dimensional, the tape is infinitely long and thus not physical.
Second, while universal, the mental model of computation is essentially matrix-vector multiplication. 
Reasoning about the output involves tracing the computation along all possible paths. A QTM has no notion of atomic instructions that combine in a predictable way, without incurring an exponential overhead for their description.

The quantum lambda calculus (QLC) is a formal language to describe computations in a quantum computer that is controlled by a classical computer.  It extends classical lambda calculus by incorporating quantum operations.  The program logic and control flow are entirely classical, written in terms of lambda abstractions, conditionals and function applications.  The data being manipulated can be quantum (i.e.,~qubits in superposition and entangled states) and reside in the quantum random access machine (QRAM) ~\cite{Phalak2023:qram}.
This calculus introduces quantum primitives: $new$ to create a qubit, $U$ to apply a unitary transformation to the qubit(s), $meas$ to measure a qubit.  To account for the physical constraints of quantum mechanics (e.g.,~the no-cloning theorem~\cite{Wootters1982:no-clone,Dieks1982:epr-comm}), the calculus is equipped with a linear type system that distinguishes between duplicable (classical) and non-duplicable (quantum) data, ensuring qubits are used only once.  The language defines a \textit{call-by-value reduction semantics}, where arguments are evaluated before being passed to functions.  Semantics include a probabilistic reduction model due to the probabilistic outcome of measurements.  The QLC supports higher-order functions and type inference, making it expressive enough to define complex quantum procedures like quantum teleportation.

The QTM, along with QLC~\cite{selinger2009quantum}, are the main abstractions used to study computation with quantum resources. Recent work by Guerrini et al.~\cite{guerrini2020quantum} provides strong evidence that having a well-developed theory of computational complexity does not suffice to provide a programming model capable of exploiting quantum resources effectively.
Both the QTM and QLC refer to state vectors and unitary transforms. While justifiable in the case of QTM (since assessing complexity is easily attainable by converting problems solvable by a QTM into circuits, then quantifying resources), the formalism is nevertheless ill-suited for algorithmic expressiveness. The explicit use of vectors and unitary transformations makes QLC more an exercise of lambda calculus applied to quantum circuits to formalize their associated types, but possibly nothing more in terms of other useful directions, such as determining whether quantum combinators can be found, whether classical ones still work, or whether higher-level instructions emerge from it.

Analogous to von Neumann's random access machine~\cite{von1993first}, the quantum random-access machine (QRAM)~\cite{knill1996conventions,miszczak2012random} and the quantum random-access stored program machine (QRASP) provide execution models centered on implementable mechanics of quantum computation. The QRAM suffers from the same symptom as QTM and QLC, since its construction depends on a vector state composed of individual qubits and unitary operations applied to it. It bears a stronger resemblance to the control mechanisms required to simulate the execution of a circuit---effectively, a \emph{hardware simulator}---than a collection of instructions. In the case of the QRASP, while recent work by Wang and Ying~\cite{wang2023quantum} makes interesting strides in building the connection between QTM and QRAM/QRASP reminiscent of Kaphengst's abstract program-controlled computing machine~\cite{kaphengst1959abstrakte}, it falls short again due to its construction corresponding to that of a QRAM, hence a quantum simulator that assumes fault-tolerance.

A configuration of a QRASP is a tuple $(\xi, \zeta, \mu, \ket{\psi}, x, y)$ which may or may not be a terminal configuration \cite{wang2023quantum}. $x, y$ are sequences of integers to be read from the input tape and written to the output tape, respectively. $\ket{\psi}$ is the state of all the quantum registers. $\mu$ is the state of the classical registers. $\xi, \zeta$ are our instruction counter and accumulator, respectively. QRASP makes use of superposition so that programs may be executed in parallel. The transition function of a QRASP tuple is defined by a dozen rules. The intructions available to a QRASP include branching, store values, addition, subtraction, loading constants, reading and printing. Quantum operations are also available, including CNOT, Hadamard, and T\@. Quantum measurement and the halt instruction round out the instruction set. A computation starts from an initial configuration $(0, 0, \mu_0, \ket{\psi_0}, in(x), \epsilon)$.

A QRAM is represented by a sequence of instructions \cite{wang2023quantum}. These instructions may be classical or quantum and include the Hadamard, CNOT and T gates as well as addition, subtraction, read/write, loading a constant, and quantum measurement. A QRAM configuration is also a tuple $(\xi, \mu, \ket{\psi}, x, y)$ where the symbols have the same meanings as given in the QRASP description above. Notably, QRAMs and QRASPs can simulate one another with polynomial overhead \cite{wang2023quantum}.

The Quantum Register Machine~\cite{zhang2024qrm} (QRM) encodes classical execution concepts, such as program counters and jump tables, into the the state itself. It does this by loading instructions and metadata into a quantum random-access memory module with its suite of qubit registers, and building up an uncomputation history as it branches.
The QRM model has not been validated against hardware, and the complication of implementing such data structures in a live system will be a hurdle. There is also the question of whether purely quantum conditional statements can be isolated away from influencing the execution pathways via entanglement. There are a few too many open questions about how this would work on real hardware to assess its viability as a hardware abstraction.

The quantum control machine (QCM)~\cite{yuan2024qcm} attempts to provide a machine model that is directly analogous to that provided by a classical CPU\@.  
The programmer has access to $n$ data registers R1,~\dots, R\textsubscript{$n$}, and the system maintains three additional control registers: a program counter, a branch control register, and an instruction register (Figure~\ref{fig:qcm}).
All of these hold exclusively qubits; there is no classical computation in the QCM model.  Program execution consists of sequentially executing assembly-level instructions under a machine model that follows a (classically) traditional fetch\dots execute\dots retire sequence.  Each step in that sequence causes updates to the control registers.  In addition to simple instructions that correspond directly to single-qubit gates and instructions like \texttt{add} and \texttt{mul} that may map to a large number of low-level operations, the programming model supports branches, including conditional branches that can be conditioned on user registers that exist in superposition.  The programmer is responsible for inserting \emph{reverse branches} (e.g.,~\texttt{rjne} in Figure~\ref{fig:qcm}) at all possible branch targets to ensure \emph{injectivity} (preservation of reversible computation) and assist the execution model with \emph{synchronization} (non-entanglement of control and data to avoid biasing measurement outcomes).

\subsection{Towards an abstraction hierarchy in quantum computing}


While respecting differences between classical and quantum resources, we modeled our list of criteria by identifying desirable properties for the production of programs and the design of future programming languages by analogy to those found in the von Neumann abstract machine~\cite{von1993first}. The historical relevance of this abstraction resides in its ability to simultaneously unify hardware models into a single execution layer, and at the same time trigger a Cambrian explosion in programming language paradigms and exemplars. Variety in programming languages is a positive sign of having a useful abstract machine, since what tends to distinguish a programming language from another is the collection of control flow structures, whose availability depends on what the execution model enables.

\begin{figure}[t]
    \centering
    \includegraphics[width=0.6\linewidth]{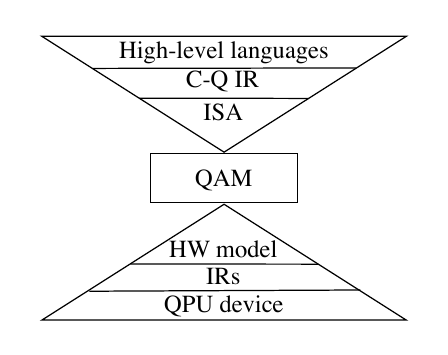}
    \caption{The quantum abstract machine is a mental model that bridges between hardware and software. It unifies the underlying hardware implementations and provides a contract atop which programming languages and frameworks can be defined.}
    \label{fig:hourglass}
\end{figure}

The dominance of quantum circuits as the primary programming model today strongly suggests we are yet to arrive at a similar tipping point. Finding a new abstract machine at the right level of expression would lead to a desired ``hourglass'' in which the multiplicity of quantum hardware platforms find a cognitively productive home (depicted in Figure~\ref{fig:hourglass}). Similarly, productive quantum programming should involve the emergence of different programming languages with specific trade-off points centered around algorithms and data structures, both of which are the main concerns of end-users and compilers.

\section{Methodology and objectives}
\label{sec:motivation}

This work was carried out according to the Design Science Research (DSR) methodology~\cite{dsr}. We decided to follow the \textit{objective-centered solution} approach, after we identified that programming quantum systems is much harder than classical due to the lack of proper abstractions. Therefore, we started the investigation by aiming at identifying what would be necessary to lower this barrier, with a set of questions: \textit{What properties of quantum abstract machines should be evaluated, and why? Do any of existing machines satisfy the resulting criteria? If not, why do they fail and, by extension, how should a new abstract machine satisfy them?}. During the \textit{Design and Development} phase we defined the criteria for a quantum abstract machine that enables proper separation between the software-intensive upper layers and hardware-oriented lower layers of the quantum software stack. Then, in the \textit{Demonstration} phase we explored the impact that the QAM fulfilling the criteria has on the major stakeholders in the quantum software stack: application programmers, system programmers and hardware implementers. For the \textit{Evaluation} phase, we revisited the existing QAMs against the criteria and position this work to related activities.

\section{Criteria for Quantum Abstract Machines}



Based on the quantum abstract models identified in the literature we extracted the following 15 criteria. Besides a description, each criterion includes examples of how an abstract machine fails to satisfy it. We consider \textit{computation with an abstract machine} to be epistemologically distinct from \textit{state evolution of a physical system}.

\criterion{Turing completeness and universality} The abstract machine can perform any finite, discrete computation any other classical or quantum machine can, regardless of performance or translation overhead. Such an abstract machine must be able to simulate other abstract machines performing finite computations of equivalent computational power. In particular, it must run any valid program without the need to alter features or properties of the definition of the machine itself. \textbf{Counterexample:} continuous-time quantum computing models with time-dependent Hamiltonians~\cite{lloyd1999quantum,kendon2020quantum} require changing physical aspects of the computational device and the specification of the machine for each specific program to execute.

\criterion{Finite symbolic state}\label{crit:sym-state} The abstract machine has a well-defined formal (symbolic) state at all times from the perspective of a programmer or compiler. The resources of the abstract machine are finite in terms of its alphabet, memory and number of states. The description of the state decomposes into addressable subunits at only one level below. The representation of the symbolic state does not equate to a recursive enumeration of the state of components of the abstract machine itself. \textbf{Counterexample:} the QTM, QRAM, and QRASP fail since these assume either an infinite tape or number of registers.

\criterion{Symbolic denotational semantics} Programmers or compilers need to reason in terms of only the symbolic state to produce correct and useful programs in the abstract machine, focusing on the results of executing symbolic programs. A program is \emph{symbolic} if (a)~it has no explicit execution or hardware model details, (b)~it does not refer to circuit elements, either classical or quantum, (c)~part of the program can be mapped to multiple equivalent functional circuit specifications without the need to know which one is implemented, and (d)~analysis of correctness of symbolic programs is independent of how each symbolic part is implemented. \textbf{Counterexample:} none of the QAMs studied here are free from the circuit model.

\criterion{Representation-independent data types} The abstract machine does not refer to circuit elements or lower circuit-based abstractions, or to specific hardware representations for its data types, i.e.,~literals and variables. In particular, the choice of number of levels available to represent data (e.g.,~bits, trits, qubits, qutrits, qudits) is fully inconsequential to the act of programming the machine. \textbf{Counterexample:} changing from qubits to qutrits or qudits changes the internals of all QAMs under study here.

\criterion{Stable instruction set architecture} All valid programs are composed of a finite and limited number of symbolic \emph{instructions}. The instruction set is compact, easy for humans to remember, and remains unchanged regardless of the evolution of hardware implementing the abstract machine. Thus, it establishes a contract for humans and compilers alike, as well as a convenient programming model. \textbf{Counterexample:} the instruction set architectures across all abstract machines depend on the choice of native gate set for quantum programming, which pertains to the hardware model.

\criterion{Verifiable formal content} The instruction set architecture defines atomic operations commonly used during problem-solving tasks that have mathematical meaning independent of their hardware realizations in terms of mappings between expected inputs and outputs. The formal content of each atomic operation corresponds to a transformation for which it is possible to specify and assert guarantees when necessary (e.g.,~admits Hoare triples~\cite{Hoare1969:hoare-logic}). \textbf{Counterexample:} continuous-variable quantum computing leads to ill-defined guarantees.

\criterion{Compact instruction representation} Instructions have a compact representation driven by their verifiable formal content regardless of the complexity of their translation into the execution or hardware models. \textbf{Counterexample}: the QTM lacks a compact representation since the entire transition table must be unfolded at all times with a combinatorial cost in the number of symbols and states.

\criterion{Classical-quantum regularity} Classical and quantum instructions live in the same abstract machine. All instructions operate at the same level of abstract representation with similar intent in their verifiable formal content. Understanding the effect of an operation of one type does not require reference to elements of either the execution or hardware model. \textbf{Counterexample:} the QRAM, QRASP and QLC have classical instructions (programming model) mixed with quantum gate execution (hardware model), hence failing this criterion.

\criterion{Degeneracy of implementation} The laws of physics allow building the abstract machine and each of its instructions. There can be at least one hardware implementation for the abstract machine, and recursively each of its components and instructions can have multiple implementations as well. All implementations preserve the formal semantics of the abstract machine and its instructions. No specific implementation choice informs the abstract properties of the programs it runs. \textbf{Counterexample:} a Zeno machine~\cite{hamkins2000infinite}, or a machine that can perform an infinite number of discrete steps in finite time, is non-physical and therefore fails to satisfy this criterion.\footnote{Surprisingly, there are impractical but theoretically possible physical implementations of abstract machines that allow using exotic relativistic effects such as the existence of a Malament-Hogarth spacetime within a rotating Kerr black hole to achieve non-Turing computation~\cite{etesi2002non}.}

\criterion{Predictable procedural composability} The properties of transformation denoted by instructions facilitate understanding of how their composition partitions possible future machine states into equivalence classes relevant to programmers or compilers. Frequent combinations of instructions may become procedures. The effect of procedures can be treated as instructions themselves at higher levels of abstraction by exposing inputs and outputs. The intent of the composition is readable from the code, and the local effect of the composition is predictable in general. \textbf{Counterexample:} QTM and QLC lack procedural composability since the computation must occur to understand its final effects; other abstract machines have complete classical procedural composability, but limited quantum procedural composability.

\criterion{Intrinsic ensemble semantics} Instructions can receive ensemble distributions as inputs and produce ensemble distributions as outputs when needed. A distribution data type over a numeric type is well defined and can be part of the signature of an instruction. Executing instructions assumes ideal hardware realizations are used, i.e.,~no ensemble semantics by virtue of classical or quantum numerical representation error. Ensemble semantics imply elision of the number of shots required to obtain a distribution using a series of instructions, since it is a detail of the execution model rather than a fundamental property of a transformation and thus irrelevant to its formal specification. \textbf{Counterexample:} none of the QRM, the QCM, or the QLC\footnote{Here we restrict ourselves to QLC as defined by~\cite{selinger2009quantum}. More sophisticated models do account for ensemble semantics, including quantum control calculi, QML, Proto-Quipper-M, categorical QLC, and probabilistic QLC.} account for distributions as outcomes.

\criterion{Resource-constructible functions} The cost of executing each instruction is quantifiable in terms of atomic, hardware-independent computational resource units (space, time, superposition, interference, entanglement). These units quantify the algorithmic complexity of algorithms for the purpose of comparison against other algorithms. \textbf{Counterexample:} continuous-time quantum computing depends on sampling rates or hand-picked time increments whose realization depends on aspects of the physical hardware implementation.

\criterion{Standard instruction cycle} The description of how the machine executes an instruction is given in terms of a single standard, mechanizable control loop involving multiple components. This control loop involves only addressable subunits as described in criterion~\ref{crit:sym-state} and does not drill down into the specifics of how subunits are constituted, their internal mechanics, or how they communicate. \textbf{Counterexample:} both the QTM and the QLC are algebraically defined in atomic steps with no decomposable access to subunits, hence lack a standard instruction cycle.

\criterion{Classical control flow} Known classical control flow structures can be implemented succinctly using only the symbolic instructions in the ISA\@. No knowledge of elements of the execution or hardware models involved in classical computing are required. \textbf{Counterexample:} the QTM lacks branching instructions entirely.

\criterion{Quantum/hybrid control flow} New purely quantum or hybrid quantum-classical control flow structures (e.g.,~sequencing) can be succinctly implemented using only the symbolic instructions in the instruction set. No elements of the execution or hardware models involved in quantum computing are required. \textbf{Counterexample:} albeit mid-circuit error detection (MCED) being a quantum control primitive,  hypothetical abstract machines with it would be disqualified due to MCED belonging to the hardware model.


\section{A desiderata for new abstract machines with quantum resources}
\label{sec-des}

The analysis above leads us to a couple of observations: (1)~existing quantum abstract machines fail to satisfy all criteria required to enable productive quantum programming at scale, and (2)~any abstract machine capable of satisfying all of them should, in principle, unify hardware models across quantum modalities while enabling new programming models we currently are unaware of at a similar level of abstraction as that present in most modern classical ones. While informative, these observations fail to provide sufficient resolution on \textit{how to find such an abstract machine}. This remains an open question, one that is both difficult and crucial to address.

We are now in a position to introduce a different, yet equally fundamental question: \textit{What should a satisfactory quantum abstract machine enable?} While indirect, this style of inquiry tends to rapidly enable both theoretical and empirical investigation toward finding the right abstract machine. These consequences of a proper execution level  abstraction constitute our first \emph{desiderata}, a list of desirable outcomes extrapolated from the criteria elaborated above. At the same time, we readily acknowledge 
that we are not able to fully predict how high-level quantum abstractions may evolve in the future.

\subsection{Desiderata for application programmers}

The most immediate consequence of having the right machine models is the ability to write classical-quantum code at the same cognitive and semantic level as the pseudocode describing the problem it intends to solve. That code should assist programmers and compilers in the choice of classical, quantum, or hybrid constructs to compactly describe the desired computation. Good abstract machines facilitate reasoning efficiently and concurrently about meaning and execution.

We expect quantum programs to integrate ensembles and their distributions seamlessly. Some instructions will naturally associate with ensemble data types, while others will represent a single value. This includes automatic type coercion to reduce the cognitive load for programmers. Quantum algorithms involve repeated execution of shots and measurements; the choice of number of shots will become immaterial to a programmer focused on how the resulting distributions transform by delegating it to the hardware level instead of making it a prominent feature of machine execution. Recent research suggests a trend toward automating the estimation of the number of shots based on properties of quantum algorithms and their underlying quantum hardware resources~\cite{miroszewski2024search,seksaria2025shots}.

We anticipate new control flow structures derived from the interplay between classical and quantum resources well beyond those already reported in the literature. These may refer directly to computed values and ensembles~\cite{yuan2024qcm} or indirectly to changes in the use of quantum resources that trigger events with usable meaning when reflecting programmer intent. Composition of classical-quantum instructions in a high-level language should result in statements and expressions at the same level of abstraction. Following past experience, the number of these new quantum-enabled control flow structures covering the space of possible and relevant programs should be compact as well.

Programmers should be able reason formally about code without worrying about its execution below the vocabulary provided by the abstract machine. Similar to how pseudocode, high-level code, and classical assembly code can be analyzed symbolically and compositionally, the new abstract machine should enable symbolic reasoning at most one degree below available control flow structures. To be productive, end-users and compilers need to verify their quantum-enabled programs fulfill certain correctness guarantees.

Performance matters for real applications. Delivering differentiated performance from a quantum processing unit (QPU) depends on effective exploitation of quantum attributes of processor resources, the essence of which is today poorly understood. However, we do know the quantum resources and attributes that need to be used effectively. The desired abstract machine should allow compilers to optimize programs by cleverly reorganizing instructions rather than having detailed physical knowledge of quantum hardware and its coupling to classical components. 
We see it as ideal that programs with quantum instructions also deliver good performance as the underlying hardware changes, but the community's understanding of what dictates strong performance appears not yet universal enough for that to be practical near-term.

Similar to how a good abstract machine enables performance through symbolic optimization, it should also enable algorithms' construction and improvement. Pseudocode should ``run'' in the mental model of execution provided by the abstract machine in the same way actual code runs on its corresponding implementation. One should be able to distinguish features of a performant algorithm from an inefficient one by means of resource-constructible functions that account for how both use time, space, superposition, interference and entanglement. In the classical world, we find reusable units of analysis depending on which control flow structures are present in the program.



A productive classical-quantum machine should help readily identify and systematically explain any advantage introduced by quantum control structures and expressions in relation to purely classical programs. This entails, when possible, the existence of recipes to transform between quantum and classical program control structures with quantification of the resulting overhead whenever these transformations are possible. For instance, transforming quantum control structures dependent upon fully superposed and/or entangled quantities is likely to result in an exponential number of classical control structures required to replicate the same outcomes.

Taking stock of all the elements above, finding the right abstract machine will likely trigger the Cambrian explosion in programming languages hinted above. These new programming languages will likely favor certain control flow structures over others, informed by the kinds of problems and applications for which they intend to be productive. Examples in the classical world include \texttt{for}, \texttt{while}, \texttt{do while} and \texttt{until} for iteration. In another case, the impossibility of unbounded iteration based on a quantum condition may be either caught and reported to the programmer, or fixed and informed as a warning during compilation.

Finally, programs written for an ISA-based architecture that contains a QPU and implements our ideal abstract machine will survive underlying hardware changes and evolutions. In particular, a good abstract machine implies that increasing the number of qubits, changing the native gate set (as long as it is universal), or selecting a different qubit modality will require no corresponding changes to be made to programs. 

\subsection{Desiderata for systems programmers}

We now place ourselves in the minds of systems programmers (e.g.,~firmware writers or compiler writers), individuals responsible for exposing emerging quantum capabilities in predictable and uniform ways.
A good abstract machine model will provide a view of the system that hardware designers can satisfy, and that programs can assume are valid and correct. In practice, this will entail the exposure of quantum resources via system interfaces with appropriate qualifiers and quantities tied to available hardware implementations. These can be used to benchmark programs quantitatively and formally correlate them with their algorithmic resource complexity.

Another consequence is that system programmers will be able to separate concerns about \textit{what functions need to be implemented} from \textit{how to execute them in detail} in a specific hardware model, an integral part of their role in the community. This happens by virtue of degeneracy of implementation. An abstract instruction can be implemented in more than one way without changing the contract it specifies. Think of the mathematical problem Shor's algorithm solves, \textit{period finding}, for which we introduce a hypothetical instruction, \texttt{qpfind}. When applied to an ensemble, it leaves an integer ``\textit{frequency}'' as a result. Systems programmers can choose to implement the Quantum Fourier Transform step using the standard recursive decomposition or classical variable extraction by Quantum Phase Estimation (QPE). The choice is immaterial for application programmers, who only expect their algorithm to run correctly and return a distribution, but not immaterial for the systems programmer who operates at the boundary that exploit specific hardware capabilities to achieve actual performance. But over time the hardware implementation of particular algorithmic patterns \textit{will} change, and it is important that when they do, programs opaquely gain the benefits of that with no, or minimal, changes.

\subsection{Desiderata for QPU manufacturers}

An effective abstract machine will highlight which quantum hardware improvements have the most impact. This reduces uncertainty on multiple fronts. First, it provides a stable contract to satisfy in which individual hardware improvements do not threaten the validity of prior investments. Second, instructions in the contract will simplify reasoning about the correctness of an implementation by providing high-level validation recipes, traceable to the hardware model and below as needed. Third, it enables innovation by showcasing specific hardware technology advantages without transferring the cost to end users. Improvements in materials, technologies, native gate sets, and quantum error-correction codes will produce value as a function of how performant QPUs are, not as a function of which codes can be run on them.

Design decisions do cascade upward across the stack, impacting compilers. RISC and CISC present paradigmatic examples from the classical world. RISC architectures favor a compact ISA with a simpler silicon implementation while translating the cost to systems programmers and compiler writers; CISC inverts this. We foresee new avenues of classical-quantum computer architecture and organization design will emerge once an abstract machine provides a contract solid enough to follow.

\section{Discussion}


\subsection{Comparison of existing QAM proposals to our criteria}

\begin{table*}
  \centering
  \caption{Analysis of prevailing quantum abstract machines}
  \label{tab:qam_comparison}
  \begin{tabular}{@{}rl*6{c}@{}}
    \toprule
    Criterion & Description & QTM~\cite{wang2023quantum} & QRAM~\cite{wang2023quantum} &  QRASP~\cite{wang2023quantum} &  QRM~\cite{zhang2024qrm} & QCM~\cite{yuan2024qcm} & QLC~\cite{valiron2024quantum} \\
    \midrule
    1 & Turing-complete \& universal & \checkmark  & \checkmark & \checkmark & \checkmark & \checkmark & \checkmark \\
    2 & Finite symbolic state & \xmark  & \xmark & \xmark & \checkmark & \checkmark & \xmark \\
    3 & Symbolic denotational semantics & \xmark  & \xmark &  \xmark & \xmark & \xmark  & \xmark \\
    4 & Representation-independent data types & \xmark  & \xmark & \xmark & \xmark & \xmark & \xmark \\
    5 & Stable instruction set architecture & \xmark  & \xmark & \xmark & \xmark & \xmark & \xmark \\
    6 & Verifiable formal content & \checkmark  & \checkmark & \checkmark & \checkmark & \checkmark & \checkmark \\
    7 & Classical-quantum regularity & \checkmark  & \xmark & \xmark  &  \checkmark${}^{\dagger}$ & \checkmark${}^{\dagger}$ & \xmark \\
    8 & Compact instruction representation & \xmark  & \checkmark & \checkmark & \checkmark & \checkmark & \checkmark \\
    9 & Degeneracy of implementation & \checkmark  & \checkmark & \checkmark & \checkmark & \checkmark & \checkmark \\
    10 &  Predictable procedural composability & \xmark  & \checkmark & \checkmark & \checkmark & \checkmark & \xmark \\
    11 & Intrinsic ensemble semantics &  \checkmark & \checkmark & \checkmark & \xmark & \xmark & \xmark \\
    12 & Resource-constructible functions & \checkmark  & \checkmark & \checkmark & \checkmark & \checkmark & \checkmark \\
    13  & Standard instruction cycle & \xmark  & \checkmark & \checkmark &  \checkmark & \checkmark & \xmark \\
    14 & Classical control flow & \xmark & \xmark & \checkmark & \checkmark & \checkmark & \checkmark \\
    15 & Quantum/hybrid control flow & \xmark & \xmark & \xmark & \checkmark & \checkmark &  \xmark \\
    \addlinespace
    Total & & 6\checkmark & 8\checkmark & 9\checkmark & 11\checkmark & 11\checkmark & 6\checkmark \\
    \bottomrule

    \multicolumn{8}{l}{%
      \footnotesize
      \rule{0pt}{3ex}%
      ${}^{\dagger}$~partial satisfaction due to explicit mention of unitary gates%
    }
  \end{tabular}
\end{table*}

In this paper we recounted how strong abstract machines drove classical computing advancement, defined fifteen criteria we believe a quantum abstract machine should satisfy to be equally effective in quantum computing, and then evaluated six quantum abstract machines against them. Two of our evaluated abstract machines --QRM and QCM-- satisfy eleven criteria; QRASP nine, QRAM eight; and QTM and QLC satisfy six.

Table~\ref{tab:qam_comparison} reveals that the major gaps are in criteria~3--5 and to a lesser extent, criterion~2. Failure to satisfy these criteria stems from a lack of separation between high-level symbolic elements and low-level hardware details. To put this more bluntly, the quantum-computing community does not yet have a good means of reasoning about quantum programs in a purely abstract sense, i.e.,~without also reasoning about how they act on quantum states. Various QAMs reference gates in their instructions and states in their definition. Changes in the hardware, such as switching from qubits to qutrits, or choosing a different universal gate set, will break the contract. While one can argue that in the latter case a compiler can simply make a one-to-one translation between gates from two different sets, the choice may have consequences in other parts of the software stack (e.g.,~choice of QECC). Better abstractions are needed, but we do not yet know what those are.

As a practical consideration, any QAM looking for wide adoption must also provide programming libraries that support easy integration into existing workflows. The quantum intermediate representation (QIR)~\cite{QIR-Alliance} is an example of this. QIR is based on the intermediate representation (IR) provided by LLVM~\cite{Lattner2004:llvm}, a widely-used and mature compilation framework. QIR adds quantum operations alongside LLVM's classical ones, inheriting all its classical representation power, ecosystem and tooling. This drastically improves the speed which QIR-supporting applications can be built and lowers the barrier of integration and support. Although QIR has no defined abstract machine and thus is not able to be directly comparable, the way it integrates into existing systems is worth highlighting.

With the lack of a quantum abstract machine satisfying all criteria and the limited programmatic implementations of proposed QAMs, we conclude that despite progress towards a proper quantum abstract machine there is not yet one that can act as a bedrock in the same way that classical abstract machines do. Whether a new machine is required or an existing one can be evolved to such a state is unclear at this point. But what \emph{is} clear is that it will take a concerted effort across the community as a whole, and across multiple disciplines, to design a quantum abstract machine that works for everyone. 

\subsection{Challenges in bridging the theory and practice}

Following past experience from the evolution of classical computers, we found convincing evidence that enabling truly productive quantum programming requires a new abstract machine that defines the execution model when quantum resources are present. More specifically, the challenge involves instructions that can be reasoned about symbolically and atomically, whose execution can be traced using the same mental model, and which can be used by programmers to build higher-level languages without knowing---and, more importantly, \emph{without having to know}---any details about the underlying hardware implementation. Existing quantum abstract machines provide part of the solution, yet remain insufficient in their ability to satisfy current and future needs.

Solving this problem, however, is a two-pronged task. Pragmatically, we need to systematically dissect existing quantum programming models, languages, and algorithm implementations and organize the resulting parts into layers. The surgical process of separating responsibilities between hardware and software is as unavoidable today for quantum computing as it was in the early days of classical computing, and will remain a hard-contact sport between hardware and software co-design for a while. In the meantime, experimenting with new control flow structures, writing code that translates quantum hardware execution to symbolic machine states, and other similar projects will help map the landscape of obstacles and opportunities. Intense experimentation with programming languages and real hardware produces two kinds of insights: we gain information about the parameters that determine what good design (i.e.,~productive expressiveness) looks like, and we collect exemplars of code from which then quantum algorithmic motifs can be extracted.

Theoretically, the challenge is much steeper. Much has been gained in the last three decades on quantum abstract machine designs, most of which may be directly reusable as a platform to stand on.
At the same time, known tools and methods seem to hit a barrier as long as we continue to insist on state vectors and unitary transformations as the canonical entities of interest.
The increasing sophistication of methods in higher algebra and related disciplines, as well as the possible interpretation of quantum computation within the framework of quantum stochastic processes---much in the spirit of~\cite{asser1959turing}---may provide a suitable starting point. Looking back at lessons learned at the intersection of theory of computation and pragmatic programming language research in the past~\cite{elgot1964random}, close collaboration will be essential.

Finding an appropriate quantum abstract machine formulation belongs to Pasteur's quadrant~\cite{stokes2011pasteur}, that of use-inspired basic research. Solving it likely entails a high failure rate, pointing to a strategy where many small teams compete and collaborate, thus covering a larger surface of attack. It is also an inevitable problem if we are to match the increasing sophistication and scale of coupled classical-quantum computers to the expected productivity of application users and programmers when using these systems. The problems raised here are both exciting and relevant, and---as Feynman said about simulating physics with computers---``by golly it's a wonderful problem, because it doesn't look so easy''~\cite{feynman2018simulating}.

\subsection{Threats to validity}

The threats to the validity of our study are discussed following the categorization provided by Wholin et al.~\cite{Wohlin2012}, dividing the validity evaluation into the following three areas.

\textit{External validity.} Existing industry efforts to define quantum specific intermediate representations (e.g.,~QIR), could bypass the QAM as a necessary layer in the implementation of the quantum software stack. Mitigation: the authors plan to engage with the relevant communities to emphasize the affordances enabled by the QAM that satisfies the criteria proposed here.

\textit{Construct validity.} Our criteria may be incomplete, failing to capture all the aspects required for a QAM to effectively serve as a separation layer between the hardware-oriented and software-intensive parts of the stack. Mitigation: we will continue to monitor emerging developments and revise the criteria as necessary.

\textit{Conclusion validity.} Quantum computing may ultimately fail to provide advantages over classical alternatives. A niche accelerator for only a few specific applications would not need a sophisticated software stack. Mitigation: unable to mitigate directly; we hope that the experimentation enabled by our layered architecture --centered around the QAM-- will help demonstrate that quantum advantage can be valuable across a broad range of application domains, mirrored by a multitude of quantum programming languages and tools that might be generic or specific to these application domains.

\section{Conclusions}

Effective and accessible abstraction hierarchies have been crucial in classical computing for enabling broad adoption and practical programming across disciplines. Achieving a similar hierarchy for quantum computing remains a major challenge, complicated by a diverse ecosystem of qubit implementation alternatives but also by rapid technological change. Central to this effort is the quantum abstract machine (QAM), which serves as the key separation layer between a programmer's mental model and the underlying quantum hardware. By proposing an evaluative criteria, we aim to guide the development of future QAMs that not only are easier to implement across diverse quantum hardware platforms but also provide a stable, effective abstraction layer for quantum programmers. This is a call to action but also one we will be taking part in. A future paper will outline precisely the aspects we would want to see in such a bedrock quantum abstract machine and its instruction set architecture.

\section*{Acknowledgment}

The authors thank Denny Dahl, Jamie Friel, Tim Mattson, Steve Reinhardt, Michał Stęchły, and Matan Vax for useful discussions.

The work presented in this paper has been funded by NSERC, the Canada Research Chairs program, UBC, NCSA, IQUIST, the IBM-Illinois Discovery Accelerator Institute, and Business Finland.  The work was also supported by the U.S. Department of Energy through Los Alamos National Laboratory (LANL)\@. LANL is operated by Triad National Security, LLC for the NNSA under contract no.~89233218CNA000001.

\clearpage

\bibliographystyle{ieeetr}
\bibliography{references}

\begin{thebibliography}{10}

\bibitem{di2024abstraction}
O.~Di~Matteo, S.~Núñez-Corrales, M.~Stęchły, S.~P. Reinhardt, and T.~Mattson, ``An abstraction hierarchy toward productive quantum programming,'' in {\em 2024 IEEE International Conference on Quantum Computing and Engineering (QCE)}, vol.~01, pp.~979--989, 2024.

\bibitem{miller1956magical}
G.~A. Miller, ``The magical number seven, plus or minus two: Some limits on our capacity for processing information.,'' {\em Psychological review}, vol.~63, no.~2, p.~81, 1956.

\bibitem{von1993first}
J.~Von~Neumann, ``First draft of a report on the {EDVAC},'' {\em IEEE Annals of the History of Computing}, vol.~15, no.~4, pp.~27--75, 1993.

\bibitem{abelson1996structure}
H.~Abelson and G.~J. Sussman, {\em Structure and Interpretation of Computer Programs}.
\newblock The MIT Press, 1996.

\bibitem{gherardi2011alan}
G.~Gherardi, ``{A}lan {T}uring and the foundations of computable analysis,'' {\em Bulletin of Symbolic Logic}, vol.~17, no.~3, pp.~394--430, 2011.

\bibitem{church1932set}
A.~Church, ``A set of postulates for the foundation of logic,'' {\em Annals of mathematics}, vol.~33, no.~2, pp.~346--366, 1932.

\bibitem{shannon1938symbolic}
C.~E. Shannon, ``A symbolic analysis of relay and switching circuits,'' {\em Electrical Engineering}, vol.~57, no.~12, pp.~713--723, 1938.

\bibitem{burks1946preliminary}
A.~W. Burks, H.~H. Goldstine, and J.~Von~Neumann, ``Preliminary discussion of the logical design of an electronic computer instrument,'' tech. rep., Institute for Advanced Studies, 1946.

\bibitem{jackson2002jsp}
M.~Jackson, ``{JSP} in perspective,'' {\em Software Pioneers: Contributions to Software Engineering}, pp.~480--493, 2002.

\bibitem{asser1959turing}
G.~Asser, ``{T}uring-{M}aschinen und {M}arkowsche {A}lgorithmen,'' {\em Mathematical Logic Quarterly}, vol.~5, no.~14-24, pp.~346--365, 1959.

\bibitem{kaphengst1959abstrakte}
H.~Kaphengst, ``Eine abstrakte programmgesteuerte {R}echenmaschine,'' {\em Mathematical Logic Quarterly}, vol.~5, no.~14-24, pp.~366--379, 1959.

\bibitem{backus1959syntax}
J.~W. Backus, ``The syntax and the semantics of the proposed international algebraic language of the {Z}urich {ACM-GAMM} conference,'' in {\em ICIP Proceedings}, pp.~125--132, 1959.

\bibitem{backus1960report}
J.~W. Backus, F.~L. Bauer, J.~Green, C.~Katz, J.~McCarthy, A.~J. Perlis, H.~Rutishauser, K.~Samelson, B.~Vauquois, J.~H. Wegstein, {\em et~al.}, ``Report on the algorithmic language {ALGOL}~60,'' {\em Communications of the ACM}, vol.~3, no.~5, pp.~299--311, 1960.

\bibitem{elgot1964random}
C.~C. Elgot and A.~Robinson, ``Random-access stored-program machines, an approach to programming languages,'' {\em Journal of the ACM (JACM)}, vol.~11, no.~4, pp.~365--399, 1964.

\bibitem{bernstein1993quantum}
E.~Bernstein and U.~Vazirani, ``Quantum complexity theory,'' in {\em Proceedings of the Twenty-Fifth Annual ACM Symposium on Theory of Computing}, pp.~11--20, 1993.

\bibitem{knill1996conventions}
E.~Knill, ``Conventions for quantum pseudocode,'' Tech. Rep. LA-UR-96-2724, Los Alamos National Laboratory, Los Alamos, New Mexico, USA, June 1996.

\bibitem{miszczak2012random}
J.~A. Miszczak, ``Random access machines,'' in {\em High Level Structures for Quantum Computing}, pp.~45--51, Springer, 2012.

\bibitem{selinger2009quantum}
P.~Selinger, B.~Valiron, {\em et~al.}, ``Quantum lambda calculus,'' {\em Semantic techniques in quantum computation}, pp.~135--172, 2009.

\bibitem{wang2023quantum}
Q.~Wang and M.~Ying, ``Quantum random access stored-program machines,'' {\em Journal of Computer and System Sciences}, vol.~131, pp.~13--63, 2023.

\bibitem{zhang2024qrm}
Z.~Zhang and M.~Ying, ``Quantum register machine: Efficient implementation of quantum recursive programs,'' 2024.

\bibitem{yuan2024qcm}
C.~Yuan, A.~Villanyi, and M.~Carbin, ``Quantum control machine: The limits of control flow in quantum programming,'' {\em Proceedings of the ACM on Programming Languages}, vol.~8, p.~1–28, Apr. 2024.

\bibitem{deutsch1985quantum}
D.~Deutsch, ``Quantum theory, the {C}hurch--{T}uring principle and the universal quantum computer,'' {\em Proceedings of the Royal Society of London. A. Mathematical and Physical Sciences}, vol.~400, no.~1818, pp.~97--117, 1985.

\bibitem{deutsch1989qcnetworks}
D.~Deutsch, ``Quantum computational networks,'' {\em Proceedings of the Royal Society of London. Series A, Mathematical and Physical Sciences}, vol.~425, no.~1868, pp.~73--90, 1989.

\bibitem{divincenzo1995two}
D.~P. DiVincenzo, ``Two-bit gates are universal for quantum computation,'' {\em Physical Review A}, vol.~51, no.~2, p.~1015, 1995.

\bibitem{Phalak2023:qram}
K.~Phalak, A.~Chatterjee, and S.~Ghosh, ``Quantum random access memory for dummies,'' {\em Sensors}, vol.~23, no.~17, 2023.

\bibitem{Wootters1982:no-clone}
W.~K. Wootters and W.~H. Zurek, ``A single quantum cannot be cloned,'' {\em Nature}, vol.~299, pp.~802--803, Oct.~28, 1982.

\bibitem{Dieks1982:epr-comm}
D.~Dieks, ``Communication by {EPR} devices,'' {\em Physics Letters A}, vol.~92, pp.~271--272, Nov.~22, 1982.

\bibitem{guerrini2020quantum}
S.~Guerrini, S.~Martini, and A.~Masini, ``Quantum {T}uring machines: Computations and measurements,'' {\em Applied Sciences}, vol.~10, no.~16, p.~5551, 2020.

\bibitem{dsr}
M.~A.~R. Ken~Peffers, Tuure~Tuunanen and S.~Chatterjee, ``A design science research methodology for information systems research,'' {\em Journal of Management Information Systems}, vol.~24, no.~3, pp.~45--77, 2007.

\bibitem{lloyd1999quantum}
S.~Lloyd and S.~L. Braunstein, ``Quantum computation over continuous variables,'' {\em Physical Review Letters}, vol.~82, no.~8, p.~1784, 1999.

\bibitem{kendon2020quantum}
V.~Kendon, ``Quantum computing using continuous-time evolution,'' {\em Interface focus}, vol.~10, no.~6, p.~20190143, 2020.

\bibitem{Hoare1969:hoare-logic}
C.~A.~R. Hoare, ``An axiomatic basis for computer programming,'' {\em Communications of the ACM}, vol.~12, pp.~576--580, Oct. 1969.

\bibitem{hamkins2000infinite}
J.~D. Hamkins and A.~Lewis, ``Infinite time {T}uring machines,'' {\em The Journal of Symbolic Logic}, vol.~65, no.~2, pp.~567--604, 2000.

\bibitem{etesi2002non}
G.~Etesi and I.~N{\'e}meti, ``Non-{T}uring computations via {M}alament--{H}ogarth space-times,'' {\em International Journal of Theoretical Physics}, vol.~41, pp.~341--370, 2002.

\bibitem{miroszewski2024search}
A.~Miroszewski, M.~F. Asiani, J.~Mielczarek, B.~L. Saux, and J.~Nalepa, ``In search of quantum advantage: Estimating the number of shots in quantum kernel methods,'' {\em arXiv preprint arXiv:2407.15776}, 2024.

\bibitem{seksaria2025shots}
M.~Seksaria and A.~Prabhakar, ``Shots and variance on noisy quantum circuits,'' {\em arXiv preprint arXiv:2501.03194}, 2025.

\bibitem{valiron2024quantum}
B.~Valiron, ``On quantum programming languages,'' {\em arXiv preprint arXiv:2410.13337}, 2024.

\bibitem{QIR-Alliance}
``{QIR} {A}alliance.'' \url{https://www.qir-alliance.org/resources/}, 2022.

\bibitem{Lattner2004:llvm}
C.~Lattner and V.~Adve, ``{LLVM}: A compilation framework for lifelong program analysis transformation,'' in {\em 2nd IEEE/ACM International Symposium on Code Generation and Optimization (CGO~2004)}, (San Jos{\'e}, California, USA), pp.~75--86, Mar.~20--24, 2004.

\bibitem{stokes2011pasteur}
D.~E. Stokes, {\em Pasteur's Quadrant: Basic Science and Technological Innovation}.
\newblock Brookings Institution Press, 2011.

\bibitem{feynman2018simulating}
R.~P. Feynman, ``Simulating physics with computers,'' in {\em Feynman and Computation}, pp.~133--153, CRC Press, 2018.

\bibitem{Wohlin2012}
C.~Wohlin, P.~Runeson, M.~H{\"o}st, M.~C. Ohlsson, B.~Regnell, and A.~Wessl{\'e}n, {\em Planning}, pp.~89--116.
\newblock Berlin, Heidelberg: Springer Berlin Heidelberg, 2012.

\end{thebibliography}

\end{document}